\documentclass[8pt,aps,showpacs,nofootinbib, twocolumn,preprintnumbers]{revtex4-1}

\usepackage{hyperref}
\usepackage{latexsym}
\usepackage{epstopdf}
\usepackage{epsf}
\usepackage{amssymb,amsmath,amsfonts,amsxtra,amsthm}
\usepackage{bm}
\usepackage{dcolumn}
\usepackage{array}
\usepackage{tabularx}
\usepackage{enumerate}
\usepackage{hhline}
\usepackage{subfigure}
\newcommand{\beq}{\begin{equation}}
\newcommand{\eeq}{\end{equation}}
\newcommand{\beqa}{\begin{eqnarray}}
\newcommand{\eeqa}{\end{eqnarray}}
\newcommand{\barr}{\begin{array}}
\newcommand{\earr}{\end{array}}

\begin{document}

\title{Inflation with teleparallelism:
Can torsion generate primordial fluctuations without local Lorentz symmetry?}
\date{\today}

\author{Yi-Peng Wu$^1$}
\affiliation{Institute of Physics, Academia Sinica,
Taipei 11529, Taiwan}

\email[Electronic address: ]{ypwu@resceu.s.u-tokyo.ac.jp}

\begin{abstract}
Arbitrary generalization to the teleparallel equivalent of general relativity loses local Lorentz invariance 
to reparametrize the orthonormal coordinate system and gives rise to asymmetry field equations.
We investigate consequences of local Lorentz violation to 
primordial fluctuations in extended single field inflationary models based on 
the scalar-tensor formulation of the torsion scalar $T$ that effectively includes $f(T)$ gravity as a special case. 
We show that despite some asymmetry part of the field equations 
are removed in a spatially homogeneous and isotropic cosmic background, 
no subhorizon scalar-perturbation mode can survive by the time of horizon crossing. 
As a result, any scalar field mediated in torsion cannot generate enough primordial density inhomogeneity alone,
even if it brings some de Sitter background solutions in generalized teleparallel gravity.
\end{abstract}

\preprint{RESCEU-29/16}

\pacs{04.50.Kd, 98.80.Jk}
\footnotetext{Current address: Research Center for the Early Universe (RESCEU),
	Graduate School of Science, The University of Tokyo, Tokyo 113-0033, Japan}

\maketitle
\section{Introduction}

The generation of small density inhomogeneity in the primoridal Universe 
is one of the most important predictions of cosmic inflation. 
In the single field inflationary paradigm one considers the inflaton to be
a canonical scalar field and quantum fluctuations in the inflaton field 
may convert to classical density perturbations after crossing the Hubble horizon.
These inflaton fluctuations simultaneously activate scalar-mode
perturbations in the gravitational action, usually addressed by `` curvature perturbations''
in the language of Einstein's general relativity (GR).
The gauge invariant observable $\zeta$ for curvature perturbations is so far
in well agreement with a Gaussian and nearly but not exactly flat spectrum
\cite{Ade:2015xua}.

Expanding the full action of the inflationary model in terms of small
perturbations is a powerful approach for computing the spectrum 
or higher-order correlation functions of $\zeta$ \cite{Maldacena:2002vr}. 
In this approach it is convenient to perform a spacetime splitting with
respect to a unit time-like vector $n_\mu = (-N, \mathbf{0})$, where
components of $n_\mu$ is determined by the metric in 
the Arnowitt-Deser-Misner (ADM) formalism \cite{ADM:1962}:
\begin{equation}\label{ADM metric}
ds^2 = - N^2 dt^2 + h_{ij}(dx^i+ N^i dt)(dx^j + N^j dt).
\end{equation}
When choosing the uniform field gauge of the inflaton, one can reparametrize
the induced metric of the spatial hypersurface as 
$h_{ij} = a^2(e^{2\zeta}\delta_{ij} + \gamma_{ij})$, where $a(t)$ is the scale factor
of the homogeneous background and $\gamma_{ij}$ parametrizes the transverse-traceless tensor fluctuation.  
Both $N$ and $N^i$ are treated as Lagrange multipliers and 
can be solved in terms of $\zeta$ and its time derivative
from the hamiltonian and momentum constraints. 

It shall be noticed that the ADM decomposition \eqref{ADM metric} assumes
no particular spacetime connection that defines
the rules of parallel transformations of the spacetime vectors. 
From this perspective similar attempts have been made in \cite{Wu:2011kh} to compute 
the primordial fluctuations 
in single field inflationary models based on Einstein's other description
of GR constructed under teleparallelism \cite{Einstein:1929} 
(see also \cite{Cho:1975dh,Hayashi:1979qx,Maluf:1994,Aldrovandi:2012}).
In the teleparallel description of gravity, the dynamical variables are living
inside the vierbein field instead of the metric, and the spacetime curvature tensor vanishes mandatorily
so that only ``torsion perturbations'' can be invoked by the inflaton field.

Driving inflation without an explicit scalar field strongly motivates 
the study on modified Lagrangian density of gravity.
In the curvature version of GR, zero-momentum de Sitter solutions are found to exist in the nonlinear extension of the 
Einstein-Hilbert action \cite{Starobinsky:1980,Starobinsky:1987zz}, which is now subject to a specific $f(R)$ gravity 
(see \cite{Sotiriou:2008rp,DeFelice:2010aj} for a review). 
It is well-known that $f(R)$ gravity can be cast into some particular types of the
Brans-Dicke theory \cite{Brans:1961sx,Clifton:2011jh}, where one may identify an auxiliary scalar field
(also called ``scalaron'' \cite{Starobinsky:1980}) to realize inflation and generate primordial fluctuations.
By virtue of the scalar-tensor correspondence \cite{Sotiriou:2008rp}, 
we will refer $f(R)$ inflation to the extended single field inflationary scenario \cite{La:1989za}.

On the other hand, de Sitter background solutions are found to exist in 
the nonlinear modification to the teleparallel equivalent of GR \cite{Ferraro:2006jd,Fiorini:2013kba,Rezazadeh:2015dza}
(or simply $f(T)$ gravity, see \cite{Cai:2015emx} and references therein) and in the extended single field model with a nonminimal coupling to
the torsion scalar $T$ \cite{Xu:2012jf,Jarv:2015odu}.
Similarly, $f(T)$ gravity is recast into a specific class of scalar-tensor theory with respect to the torsion
scalar \cite{Sotiriou:2010mv}, such that we can study the inflationary fluctuations in a unified manner based on 
the general scalar-tensor setup within teleparallelism.
In this work we aim to recognize whether the torsion induced scalaron is a suitable candidate for inflaton.
We improve the parametrization for vierbein variables in \cite{Wu:2011kh}
and perform a research independent to the metric perturbation approach \cite{Cai:2011tc,Rezazadeh:2015dza}.
 
Due to the presence of local Lorentz violation in nonlinear teleparallelism \cite{Li:2010cg,Sotiriou:2010mv},
generalized teleparallel gravity is essentially unhealthy unless closely recovers the limit of GR \cite{Ong:2013qja,Izumi:2013dca,Chen:2014qtl}.
Even if a simple cosmic background is assumed, 
local Lorentz symmetry breaking will deny the coordinate reparametrization in the orthonormal frame from removing some dynamical variables, 
and thus additional field equations must arise to account for those non-vanished degrees of freedom. 
However, consequences of additionally induced field equations in cosmological perturbations were not addressed in previous study
\cite{Chen:2010va,Dent:2011zz,Cai:2011tc,Wu:2011kh,Izumi:2012qj,Rezazadeh:2015dza}.
In this work, we scrutinize implications of these unusual equations of motion. 
We show that the asymmetry part of the field equations, once exists, always suppress the generation of the gauge invariant
perturbation $\zeta$ in the general scalar-tensor formulation of teleparallel gravity.

\section{teleparallel inflation}

Teleparallel gravity naturally defines vectors and tensors
in the orthonormal frame \cite{Indices} with bases $\hat{e}_A$ satisfying 
$\hat{e}_A \cdot \hat{e}_B = \eta_{AB}$, where 
$\eta_{AB} = \text{diag}(-1,1,1,1)$. 
The orthonormal bases are related to the general coordinate bases through the
vierbein field $e_A^{\;\;\mu}$ as $\hat{e}_A = e_A^{\;\;\mu}\partial_\mu$.
The spacetime metric 
\begin{equation}
\label{metric-vierbein relation}
g_{\mu\nu} = \eta_{AB}e^A_{\;\;\mu}e^B_{\;\;\nu}
\end{equation}
is given by the dual vierbein $e^A_{\;\;\mu}$, where 
$e^A_{\;\;\mu}e_A^{\;\;\nu} = \delta_\mu^\nu$ and
$e_A^{\;\;\mu}e^B_{\;\;\mu} = \delta_A^B$.
The absolute parallel (zero-curvature) condition $\nabla_\mu e^A_{\;\;\nu} = 0$  
 is achieved by the Weitzenb\"{o}ck connection
$\Gamma^\lambda_{\;\;\nu\mu} = e_A^{\;\;\lambda}\partial_\mu e^A_{\;\;\nu}$.
The Weitzenb\"{o}ck connection is related to the Levi-Civita connection
$\bar{\Gamma}^\lambda_{\;\mu\nu}$ (the spacetime connection in GR) via
\begin{equation}\label{W-connection decomposition}
\Gamma^\lambda_{\;\mu\nu} = 
\bar{\Gamma}^\lambda_{\;\mu\nu} + K^\lambda_{\;\mu\nu},
\end{equation}
where 
$K^\lambda_{\;\mu\nu} = \frac{1}{2}(T^{\;\;\lambda}_{\mu\;\;\nu} + 
T^{\;\;\lambda}_{\nu\;\;\mu} - T^\lambda_{\;\;\mu\nu})$
is the contorsion and
\begin{equation}
T^\lambda_{\;\;\mu\nu} = 
\Gamma^\lambda_{\;\;\nu\mu} -\Gamma^\lambda_{\;\;\mu\nu}
\end{equation}
is the torsion tensor in teleparallel gravity. 

We expect the standard inflationary perturbations \cite{Maldacena:2002vr}
to be recovered in the teleparallel equivalent formulation of GR
with the gravitational Lagrangian given by the torsion scalar
\cite{Einstein:1929, Maluf:1994}
\begin{equation}
T = S_\lambda^{\;\;\mu\nu} T^\lambda_{\;\;\mu\nu},
\end{equation}
where $S_\lambda^{\;\;\mu\nu} = 
\frac{1}{2}(K^{\mu\nu}_{\;\;\;\;\lambda} + 
\delta^\mu_\lambda T^{\alpha\nu}_{\;\;\;\;\alpha} -
\delta^\nu_\lambda T^{\alpha\mu}_{\;\;\;\;\alpha})$.
Taking the connection decomposition \eqref{W-connection decomposition} 
into the torsion scalar,
one finds $T = -\bar{R} - 2\bar{\nabla}_\mu T^{\alpha\mu}_{\;\;\;\;\alpha}$,
where $\bar{R}$ is nothing but the curvature scalar in GR and 
$\bar{\nabla}_\mu$ is the covariant derivative with respect to the
Levi-Civita connection. Therefore the action 
$S = \int d^4x\, e\, T$ manifests the
Einstein-Hilbert action upto a total divergence, where $e = \sqrt{-g}$.


To retain a general discussion, we consider the action based on the scalar-tensor formulation 
with respect to the torsion scalar as \cite{Izumi:2013dca,Jarv:2015odu}:
\begin{equation}\label{inflation model}
S=\int d^4 x e \left[ \frac{F(\phi)}{2} T + 
\frac{Z(\phi)}{2}\partial_\mu \phi\partial^\mu\phi + V(\phi)\right] ,
\end{equation}
where $\phi$ shall serve as the inflaton field. 
Non-trivial coupling between torsion and a scalar field is also seen by the higher-dimensional teleparallel gravity \cite{Geng:2014yya,Geng:2014nfa}.
The action \eqref{inflation model} can mimic the dynamics of $f(T)$ gravity according to
the reformulation: $F(\phi) = \phi \equiv df/dT$, $Z(\phi) = 0$ and 
$V(\phi) = f(T(\phi)) - \phi T(\phi)$, provided a sufficient condition $d^2f/dT^2\neq 0$ \cite{Sotiriou:2010mv}. 
On the other hand, the model with a nonminimal coupling to the
torsion scalar \cite{Geng:2011aj} is recovered by the choice
$F(\phi) = 1 + 2\xi\phi^2$, $Z(\phi) = 1$.

We express the field equations in the general coordinate bases, and the
components are given by
\begin{eqnarray}\label{field eqs}
F(\phi) & G_{\mu\nu} &\; - \; 
2 F_\phi S_{\mu\nu}^{\;\;\;\;\lambda}\partial_\lambda \phi  
 \\ 
&=&g_{\mu\nu} \left[ \frac{Z(\phi)}{2}(\partial\phi)^2 + V(\phi)\right] - Z(\phi)\partial_\mu\phi \partial_\nu \phi,
\nonumber
\end{eqnarray}
where $G_{\mu\nu}$ is the Einstein tensor defined in GR and 
$F_\phi\equiv d\, F/ d\phi$. In Eq. \eqref{field eqs}, the manipulation
$G_\mu^{\;\;\nu}=e^A_{\;\mu}G_A^{\;\;\nu}$, where
$G_A^{\;\;\nu} = 2 e^{-1}\partial_\mu(ee_A^\lambda S_\lambda^{\;\mu\nu})
- 2 e_A^\lambda T^\rho_{\;\mu\lambda}S_\rho^{\;\nu\mu}
-e_A^\nu T/2$ has been used.
The equation of motion for $\phi$ is
\begin{equation}\label{EoM: phi}
Z(\phi)\bar{\square}\phi + \frac{1}{2} Z_\phi(\partial\phi)^2 
-\frac{F_\phi}{2} T - V_\phi = 0,
\end{equation}
where $\bar{\square}\equiv \bar{\nabla}^\mu\bar{\nabla}_\mu$.
 The clear evidence of the local
Lorentz violation in the action \eqref{inflation model}
is the existence of the antisymmetry part of \eqref{field eqs}
(see the discussion in \cite{Li:2010cg}), which reads
\begin{equation}\label{field eqs-anti}
F_\phi
\left( S_{\mu\nu}^{\;\;\;\;\lambda} - S_{\nu\mu}^{\;\;\;\;\lambda} \right)
\partial_\lambda \phi = 0.
\end{equation}
The presence of Eq.\eqref{field eqs-anti} indicates that the total components
of the field equations are 16, responsible to all the 16 components in the
vierbein field $e^A_{\;\mu}$. 

It is noteworthy that some of the 6 components released due to the
local Lorentz violation may not finally become physical degrees of freedom.
The reason is that, in some cases, the local Lorentz group may not
entirely broken and some degrees of freedom can be removed
by the unbroken sub-symmetry group \cite{Ferraro:2014owa}. 
This is indeed the case with the background vierbein choice 
\begin{equation}
\label{homogeneous vierbein}
e^A_{\;\;\mu} = \text{diag}(1,a,a,a),
\end{equation}
following the homogeneous and isotropic cosmological principle.
One shall keep in mind that generalized teleparallel gravity admits a sudden transition of the homogeneous and isotropic background
to any inhomogeneous or anisotropic kinds \cite{Izumi:2013dca}. 
For convenience, let us assume that such transition does not happen during the epoch of inflation.
 
With the background choice \eqref{homogeneous vierbein}, 
the antisymmetry equations \eqref{field eqs-anti} vanish
identically and the field equations \eqref{field eqs} are given by
\begin{eqnarray}
\label{Friedmann1}
3H^2 F &=& \frac{Z}{2}\dot{\phi}^2 + V, \\
\label{Friedmann2}
-(2\dot{H} + 3H^2) F &=& \frac{Z}{2}\dot{\phi}^2 - V +2 H F_\phi \dot{\phi},
\end{eqnarray}
where $T = 6H^2$. The background equation of motion for $\phi$ is 
\begin{equation}
\label{EoM in FRW}
Z\left( \ddot{\phi}+ 3H\dot{\phi}\right) + \frac{Z_\phi}{2} \dot{\phi}^2
+ 3H^2 F_\phi + V_\phi =0.
\end{equation}

It is not our goal to build up a new inflation model, but we can easily show the existence of de Sitter background solutions
in the extended single-field theroy \eqref{inflation model} (see also \cite{Ferraro:2006jd,Fiorini:2013kba,Rezazadeh:2015dza,Jarv:2015odu}). For example, let us consider $Z = c_1$ and $V = c_2 F$, 
where $c_1$ and $c_2$ are some constants. We then combine the Friedmann euqations \eqref{Friedmann1} and \eqref{Friedmann2} to obtain
\begin{eqnarray}
\label{Friedmann1_dS}
&&(3H^2 - c_2)F = \frac{c_1}{2}\dot{\phi}^2, \\
\label{Friedmann2_dS}
&&-2\dot{H}F = (c_1\dot{\phi} + 2HF_\phi)\dot{\phi}, 
\end{eqnarray}
 and the equation of motion \eqref{EoM in FRW} may be rewritten as
\begin{equation}
\label{EoM_dS}
\ddot{\phi} + 3H\dot{\phi} + \frac{F_\phi}{2F}\dot{\phi}^2 = 0.
\end{equation}
Assuming $c_2 > 0$, one simply check that $(H, \phi) = (\pm\sqrt{c_2/3}, \phi_\ast)$ with an arbitrary constant $\phi_\ast$ 
are solutions for the above equations \eqref{Friedmann1_dS}, \eqref{Friedmann2_dS} and \eqref{EoM_dS}. In particular, it is straightforward 
to perform the perturbations $H = H_\ast + \delta H$, $\phi = \phi_\ast +\delta\phi$ around the fixed points to find that 
$H_\ast = \sqrt{c_2/3}$ is a stable de Sitter solution.
Since $F(\phi)$ approaches to a constant toward the fixed point, the stable de Sitter attractor $(H, \phi) = (H_\ast, \phi_\ast)$ satisfies the viable conditions considered in \cite{Jarv:2015odu}.

\section{primordial fluctuations}

Let us now compute the cosmological perturbations by virtue of the 
ADM formalism \eqref{ADM metric}. 
To our purpose, we will parametrize the 16 variables in the vierbein field
in a separate way.
The first step is to identify a primary vierbein with respect to the homogeneous choice 
\eqref{homogeneous vierbein} that satisfies the condition \eqref{metric-vierbein relation} 
for the ADM metric \eqref{ADM metric}. A suitable solution is found in \cite{Wu:2011kh} of the form
\begin{eqnarray}
\label{ADM-vierbein 0th}\nonumber
\bar{e}^0_{\;\mu}=(N,{\bf 0})\;\; &,&\;\;\;
\bar{e}^a_{\;\mu}=(N^a,h^a_{\,\,i})\;,\\
\bar{e}^{\,\,\mu}_{0}=(1/N,-N^i /N)\;\; &,&\;\;
\bar{e}^{\,\,\mu}_{a}=(0,h_a^{\,\,i})\;,
\end{eqnarray}
where $N^a \equiv N^i h^a_{\,\,i}$.
Here $h_{ij} = \eta_{ab}h^a_{\,\,i}h^b_{\,\,j}$ manifests the induced metric
of the 3-surface, so that $h^a_{\,\,i}$ is a possible choice of the
induced vierbein (the dreibein hereafter).
The representation of $\bar{e}^A_{\;\mu}$ in \eqref{ADM-vierbein 0th}
is specially chosen such that $\bar{e}^{\,\,\mu}_{0}$ coincides with the unit vector $n^\mu = (1/N,-N^i /N)$, 
	where only dynamical variables responsible to the metric are shown. 
	In particular, one can parametrize $\zeta$ and $\gamma$ in $h^a_{\,\,i}$ to arbitrary order of interest.

The next step is to put together the components corresponding to the
local Lorentz invariance. Schematically, we have decomposed
the vierbein with respect to the infinitesimal Lorentz transformation
$\Lambda^{A}_{\;\,B}(x) = (e^{\omega})^{A}_{\;\,B} = \delta^{A}_{\;B}
+ \omega^{A}_{\;B} + \frac{1}{2}\omega^{A}_{\;C}\omega^{C}_{\;B} +...$ as 
\begin{equation}\label{vierbein expansion}
e^A_{\;\;\mu} = (e^{\omega})^{A}_{\;B} \bar{e}^B_{\;\;\mu}
 = {}^{(0)}e^A_{\;\;\mu} + {}^{(1)}e^A_{\;\;\mu} +...,
\end{equation}
where $\omega_{AB} = -\omega_{BA}$ and
 $\bar{e}^A_{\;\;\mu}$ is given by \eqref{ADM-vierbein 0th}.

If the local Lorentz symmetry is broken, then $\omega^{A}_{\;B}$ is no longer
the transformation matrix but rather than the ``Goldstone modes'' of the
symmetry breaking. We may parametrize these modes as
\begin{equation} \label{extra vierbein DoFs}
\omega^{0}_{\;B} = (0, \omega_b)\;\;\; ,\;\;\;
\omega^{a}_{\;B} = (\omega^a, B^{a}_{\; b}),
\end{equation}
where $\omega^a = \eta^{ab}\omega_b$ and $B_{ab} + B_{ba} = 0$.
We then define the spatial vector $\omega^i = \omega^a h_a^{\;i}$ 
and the antisymmetric spatial tensor $B_{ij} = B_{ab}h^a_{\;i} h^b_{\;j}$.

Since $\omega^{A}_{\;B}$ satisfies $\eta_{AB} 
= \eta_{CD}(e^{\omega})^{C}_{\;\,A}(e^{\omega})^{D}_{\;\,B}$, 
Eq. \eqref{metric-vierbein relation} implies that
$\omega^a$ and $B^{a}_{\; b}$ can not contribute to the metric
and the curvature scalar, which is given by
\begin{equation}\label{Ricci scalar decomposition}
\bar{R} = \bar{R}^{3} + \bar{\Sigma}^{ij}\bar{\Sigma}_{ij} -\bar{\Sigma}^2
- 2 \bar{\nabla}_\mu (n^\mu \bar{\Sigma}) - \frac{2}{N} \bar{\Delta} N,
\end{equation}
where $\bar{\Sigma}_{ij} = \frac{1}{2N}(\dot{h}_{ij} - 
\bar{\nabla}_i N_j - \bar{\nabla}_j N_i)$
is defined in the same way as the extrinsic curvature in GR,
while $\bar{\Delta}\equiv h^{ij}\bar{D}_i\bar{D}_j$ and $\bar{D}_i$
is the 3-covariant derivative with respect to the 3-Levi-Civita
connection ${}^{3}\bar{\Gamma}^{i}_{\;\,jk}$.

Using the projection tensor $\perp_\mu^{\;\,\nu} = h_\mu^{\;\,\nu}$,
where $h_{\mu\nu} = g_{\mu\nu} + n_\mu n_\nu$, we can obtain
the induced connection in the 3-surface through the definition 
$D_i A_j \equiv \perp_i^{\;\,\mu}\perp_j^{\;\,\nu} \nabla_\mu A_\nu$
for an arbitrary vector $A_\mu$ lies in the 3-surface ($n^\mu A_\mu =0$). 
Taking the zeroth order vierbein \eqref{ADM-vierbein 0th} into calculation,
we find that ${}^{3}\Gamma^{i}_{\;\,jk} = h_a^{\;i}\partial_k h^{a}_{\;j}$
is a 3-Weitzenb\"{o}ck connection
so that the intrinsic curvarture of the 3-surface is zero
(namely it is a teleparallel hypersurface).
Note that the zero-curvature condition of the 3-surface holds at 
any-order expansion of the vierbein with respect to $\omega^{A}_{\;B}$.
This is acheived through a redefinition of the dreibein
$\tilde{h}^{a}_{\;i}\equiv (e^B)^a_{\;b}h^{b}_{\;i}$, where
one can check that 
$h_{ij}=\eta_{ab}\tilde{h}^a_{\,\,i}\tilde{h}^b_{\,\,j} 
=\eta_{ab}h^a_{\,\,i}h^b_{\,\,j}$.

In order to compute the quadratic action, we shall consider the 
expanded vierbein \eqref{vierbein expansion} upto second order
in $\omega^{A}_{\;B}$. In terms of the parametrization 
\eqref{extra vierbein DoFs}, the first order expansion reads
\begin{eqnarray}
{}^{(1)}e^0_{\,\,\mu} = (N^a\omega_a, \omega_i)\; , \;
{}^{(1)}e^a_{\,\,\mu} = 
(N\omega^a +N^bB^a_{\;b}, B^a_{\;i}), 
\;\;\;\;\;\; \\
{}^{(1)}e_0^{\,\,\mu} = (0,\; -\omega^i)\; , \;
{}^{(1)}e_a^{\,\,\mu} = 
(-\frac{\omega_a}{N} ,\; B_a^{\;i} + \frac{N^i}{N}\omega_a).
\;\;\;\;\;\;
\end{eqnarray}
The second order expansion takes the form
\begin{eqnarray}\nonumber
{}^{(2)}e^0_{\,\,0} &=& \frac{1}{2}N \omega_a\omega^a
+ \frac{1}{2}N^b\omega_c B^c_{\,\,b}, \\\nonumber
{}^{(2)}e^0_{\,\,i} &=& \frac{1}{2} \omega_a B^a_{\,\,i}, \\\nonumber
{}^{(2)}e^a_{\,\,0} &=& \frac{1}{2} N^b\omega^a\omega_b 
+ \frac{1}{2} N B^a_{\,\,b}\omega^b + \frac{1}{2} N^b B^a_{\,\,c}B^c_{\,\,b}, \\
{}^{(2)}e^a_{\,\,i} &=& \frac{1}{2} \omega^a\omega_i 
+ \frac{1}{2} B^a_{\,\,c}B^c_{\,\,i}, \nonumber
\end{eqnarray}
where the dual field components are
\begin{eqnarray}
{}^{(2)}e_0^{\,\,0} &=& \frac{1}{2N} \omega_a\omega^a , \nonumber\\
{}^{(2)}e_0^{\,\,i} &=& -\frac{N^i}{2N} \omega_a\omega^a 
-\frac{1}{2} \omega_b B^{bi}, \nonumber\\
{}^{(2)}e_a^{\,\,0} &=& -\frac{1}{2N} B_{ab}\omega^b , \nonumber\\
{}^{(2)}e_a^{\,\,i} &=& \frac{1}{2} \omega_a\omega^i 
+ \frac{1}{2} B_{ab}B^{bi} + \frac{N^i}{2N}B_{ac}\omega^c. \nonumber
\end{eqnarray}
It is convenient to use $D_i\omega_j = \partial_i\omega_j - {}^{3}\Gamma^{l}_{\;\,ji} \omega_l = h^a_{\; j}\partial_i \omega_a$.
Together with \eqref{Ricci scalar decomposition}, we can write down the
second order action of \eqref{inflation model} as
\begin{eqnarray}\label{second order action}
{}^{(2)}S &=& \int dtd^3x\, N \sqrt{h} 
\left\lbrace  \frac{F(\phi)}{2} T \right. \\\nonumber
&+& 
\left. \frac{Z(\phi)}{2} 
[ h^{ij} \partial_i\phi\partial_j\phi - (n^\mu\partial_\mu \phi)^2 ] +  V(\phi) \right\rbrace  ,
\end{eqnarray}
where
\begin{eqnarray}
T = &-& \bar{R}^{3} + \bar{\Sigma}^2 - \bar{\Sigma}^{ij}\bar{\Sigma}_{ij} 
+ \frac{2}{N} \bar{\Delta} N  
- \frac{2}{N} \bar{D}_i (h^{ij} N T^{\alpha}_{\;\; j\alpha}) 
\nonumber
\\ \nonumber
&-& 2 \bar{\nabla}_\mu 
\left[ n^\mu D_i \omega^i + \frac{n^\mu}{N} D_i(N^b B^i_{\; b})\right] 
\\ \nonumber
&-& \bar{\nabla}_\mu 
\left[ n^\mu(B^{ij}D_j\omega_i + \omega_j D_i B^{ij})\right]+...
\end{eqnarray}
Those unlisted terms above are higher-order contributions, keeping in mind
that $N^i$ and $\partial_i N$ are at least first order perturbations.

Let us examine the first order solution of $N$ and $N^i$ in the
uniform field gauge where $\delta\phi = 0$ with the parametrization
of the dreibein $h^a_{\;i} = ae^\zeta\delta^a_{\;i}$ (the general
coordinate is now completely fixed). 
For this purpose we only need to
focus on the scalar and vector perturbations and keep $\gamma^{ij}$ and 
 the three modes in $B^{ij}$ 
 (which are pseudoscalar and pseudovector modes \cite{Izumi:2012qj}) aside. 
The resulting hamiltonian and momentum
constraints from \eqref{second order action} are 
\begin{eqnarray}
&& F(\bar{R}^3 + \bar{\Sigma}^2 - \bar{\Sigma}^{ij}\bar{\Sigma}_{ij})
- \frac{Z}{N^2}\dot{\phi}^2 - 2V =0, \\
&& \bar{D}_i (\bar{\Sigma}\delta^i_{\;j} - \bar{\Sigma}^i_{\;j}) =0.
\end{eqnarray}
Taking $N = 1 + N_1$ and $N^i = \partial^i\psi + N^i_T$ with 
$\partial_i N^i_T = 0$, we find that $N^i_T=0$ and that
\begin{equation}
\label{solutions:uniform field gauge}
N_1 = \frac{\dot{\zeta}}{H},\;\;
\psi = \frac{-\zeta}{2a^2H} + \chi, \;\; 
\partial^2\chi = \frac{Z\dot{\phi}^2\dot{\zeta}}{2H^2F}.
\end{equation}

We can now solve the equation of motion for $\omega^i$ and $B^{ij}$ 
by using the results \eqref{solutions:uniform field gauge}. 
To perform the variation with respect to
$\omega^i$ and $B^{ij}$, it is convenient to use the relation 
$D_i\omega^i = \bar{D}_i\omega^i + {}^{3}K^i_{\;ji}\omega^j$, where
${}^{3}K^i_{\;jk}$ is the 3-contorsion defined with respect to the 3-torsion 
${}^{3}T^i_{\;jk} = {}^{3}\Gamma^{i}_{\;\,kj} - {}^{3}\Gamma^{i}_{\;\,jk}$
on the 3-surface. The resulting equations from variation of the action \eqref{second order action} are
\begin{eqnarray}\label{Eom omega: uniform field}
&& {}^{3}T^j_{\;ji} F_\phi = h_a^{\;j}(\partial_j h^a_{\;i} - \partial_i h^a_{\;j}) F_\phi =0, \\
\label{eom Bij}
&&\left( \partial_j \omega_i - \partial_i \omega_j \right) F_\phi = 0,
\end{eqnarray}
at first order in the uniform field gauge. One can check that these equations 
are identical to the linearized equations of \eqref{field eqs-anti}.

If $F_\phi =0$, both Eq. \eqref{Eom omega: uniform field} and Eq. \eqref{eom Bij} vanish
as the theory converge to the GR limit.
This was already pointed out in Ref. \cite{Chen:2010va,Dent:2011zz}.
Since the kinetic term $Z(\partial\phi)^2$ can be canonicalized through 
a redefinition of $\phi$, the solutions \eqref{solutions:uniform field gauge} result in a standard quadratic action of 
the single field inflation model \cite{Maldacena:2002vr} up to a constant $F$.
In general cases where $F_\phi\neq 0$,
the decomposition $\omega^i =\partial^i \alpha + \omega^i_T$ with
$\partial_i\omega^i_T = 0$ then leads to the solution $\omega^i_T =0$
from \eqref{eom Bij}. The gradient of Eq. \eqref{Eom omega: uniform field}
gives $F_\phi\partial^2 \zeta =0 $, which implies $k^2 \zeta_k =0$, 
where $\zeta_k$ denotes a non-zero Fourier mode of $\zeta$.
Since the wavenumber $k$ is arbitrary, we find an unexpected result
$\zeta_k = 0$ for each $k \in (0, k_\ast]$, where $k_\ast$
denotes some cutoff scale well inside the horizon \cite{Sub-horizon}
(and the zero-mode solution $\zeta = \zeta_0(t)$ is always rescaled into the scale factor). 
Taking $\zeta=0$ into Eq. \eqref{EoM: phi}, one then finds that $\alpha = 0$. 
One can see an obvious discontinuity in the primordial fluctuations with the presence of an amazingly small $F_\phi$.
To smooth the solution in the limit $F_\phi\rightarrow 0$, we impose a possible choice of the cutoff as 
$k_\ast = \sqrt{\vert F_\phi \vert}\, k_{end}$, where $k_{end}$ is the horizon scale at the end of inflation.

We remark that the 3-torsion 
	${}^{3}T^i_{\;jk} = h_a^{\;j}(\partial_j h^a_{\;i} - \partial_i h^a_{\;j})$ is indeed a tensor form of the dreibein field on the 3-surface
	so that any non-diagonal background choice given by a spatial rotation
	of the dreibien will not change the form of Eq. \eqref{Eom omega: uniform field}. 
	As a result, the 3-torsion (and therefore the scalar perturbation $\zeta$) must vanish if $F_\phi \neq 0$.

 
The other convenient gauge for computation is the spatially flat slicing
in which the general coordinate is fixed such that 
$h^a_{\;i} = a\,\delta^a_{\;i}$ and $\delta\phi = \delta\phi(t,x)$.
In this gauge $N$ and $N^i$ are solved as
\begin{eqnarray}
&& N_1 = \frac{Z\dot{\phi}}{2HF}\delta\phi, \;\;\; N^i_T = 0, \\\nonumber
&& \partial^2 \psi = -\frac{Z \dot{\phi}^2}{2H^2F} \frac{d}{dt}
\left( \frac{H}{\dot{\phi}}\delta\phi\right) +
\frac{3HF_\phi}{F}\delta\phi.
\end{eqnarray}
Similarly, the linearized equation of motion for $\omega^i$ and $B^{ij}$
(or Eq. \eqref{field eqs-anti}) gives
\begin{eqnarray}\label{EoM omega: flat slicing}
&&\partial_i \left( N n^\mu \partial_\mu F\right) =0, \\
&&\left( \partial_j \omega_i - \partial_i \omega_j \right)
n^\mu \partial_\mu F = 0,
\end{eqnarray}
which indicates $\delta\phi = \delta\phi_0 (t)$,
$\omega_T^i =0$ if $F_\phi\neq 0$. Therefore, any subhorizon mode 
of $\delta\phi_k$ at the onset of inflation is eliminated when stretched toward the horizon scale unless $F_\phi =0$.

It is now clear that the gauge invariant variable
\begin{equation}
\zeta = \varphi - \frac{H}{\dot{F}}\delta F,
\end{equation}
where $\varphi$ parametrizes the diagonal scalar mode in $h_{ij}$,
is at best a homogeneous solution in either the uniform field gauge ($\delta\phi = 0$)
or in the spatially flat gauge ($\varphi =0 $) unless $F$ is a constant, due to the
constrain equation \eqref{field eqs-anti}.
Indeed, this result is consistent with observations in the Newtonian gauge 
(that is to set $\psi = 0$ and $h^a_{\;i} = a e^{\varphi}\,\delta^a_{\;i}$
in our notation)
where one obtains $\varphi = H\delta\phi/\dot{\phi}$ for non-zero modes in
 the $f(T)$ gravity \cite{Wu:2012hs} as well as 
 the model of nonminimal coupling 
 $F(\phi)= 1 + 2\xi\phi^2$ \cite{Geng:2012vn}. 

\section{Conclusions}
In this work we considered the generation of primordial fluctuations in the extended single field inflationary scenario 
based on a scalar-tensor formulation of teleparallel gravity which effectively includes $f(T)$ theory as a special case. 
Let us summarize the fate of the 16 + 1 perturbation variables reside in 
$e^A_{\;\mu}$ and $\phi$ under the idealized homogeneous and isotropic background \eqref{homogeneous vierbein}.
We always remove 4 variables via the general coordinate invariance and
we can apply the hamiltonian and momentum constraints 
to solve the other 4 variables in $N$ and $N^i$.
We have put aside the two tensor degrees of freedom for separate discussion.
In fact, by taking the linear parametrization 
$h^a_{\;i} = a(\delta^a_{\;i} +\gamma^a_{\;i}/2)$ for the tensor variable, 
one can observe that 
$\gamma_{ij}= \eta_{ab}(\delta^a_{\;i}\gamma^b_{\;j} 
+ \delta^b_{\;j}\gamma^a_{\;i})/2$ only involves in $\bar{R}^3$ and 
$\bar{\Sigma}_{ij}$ in the action \eqref{second order action}.
Therefore, the quadratic action for $\gamma$ is the identical to that of the
Einstein-Hilbert action upto the factor $F$.
 
The equation of motion \eqref{EoM: phi} helps to replace one scalar mode by the others so that there are now 6 modes left for
the 6 asymmetry field equations \eqref{field eqs-anti}. If $F$ is evolving with time,
three of the 6 remaining variables are eliminated by the asymmetry field equations \eqref{field eqs-anti}, which are 
accounted by a scalar mode (that is $\zeta$ or $\delta\phi$, depending on the gauge) and a transverse vector mode ($\omega^i_T$).
Nevertheless, owing to the background choice \eqref{homogeneous vierbein}, 
we notice that the equation of motion for the other three variables inside $B_{ij}$ never shows up. 
Further investigation beyond linear perturbation level is required, given that
those modes released by local Lorentz violation are found to be pathological in non-trivial spacetime background
\cite{Ong:2013qja,Izumi:2013dca,Chen:2014qtl}.
Therefore if there exists any higher-order coupling between $B_{ij}$ and other dynamical modes it would jeopardize 
 the theory even if the sudden background transition has been omitted, as a priori assumption.  

After clarifying all the equations of motion,
we conclude that the general teleparallel scalar-tensor theory only admits zero-momentum solutions 
for the scalaron,
and the extended single field inflationary model \eqref{inflation model} exhibits no dynamical
perturbation modes other than the tensor fluctuations unless converging to the GR limit. 
In particular, the modified Lagrangian $f(T)$ itself is not enough to explain
the horizon problem, the flatness problem and the origin of the structure of the Universe at once.
This is a drastically different prediction from the well-known result given by $f(R)$ 
(or $f(\bar{R})$ according to our notations) theories \cite{Starobinsky:1980,DeFelice:2010aj}.
To realize a suitable initial state of the Universe, one may consider to 
restore the local Lorentz symmetry to get rid of the asymmetry field equations \cite{Krssak:2015oua},
or to introduce some extra matter to trigger the scalar perturbation $\zeta$ \cite{Rezazadeh:2015dza}. 
In the former case successful primordial fluctuations will be promised by using the proper covariant formulation
of generalized teleparallel gravity \cite{Krssak:2015oua}, 
while in the latter case one would have to go beyond the single field inflationary paradigm.
We remark that any additional matter will not involve in the asymmetry field equations \eqref{field eqs-anti},
if it is protected by local Lorentz symmetry 
(as the scenario considered in \cite{Ferraro:2006jd,Fiorini:2013kba,Rezazadeh:2015dza}). 
As shows in the flat slicing, one can see that if Eq. \eqref{EoM omega: flat slicing}
is unchanged then the solution $\delta\phi_k = 0$ persists. 

Finally we note that no trivial Einstein frame of the generalized
teleparallel gravity \eqref{inflation model} is avalible through a
conformal transformation where the scalar field is fully decoupled
with torsion. For instance, by taking 
$\hat{e}^A_{\;\mu} = \sqrt{F}e^A_{\;\mu}$ for the vierbein, 
we can reach a minimally coupled torsion scalar $\hat{T}$ in the new frame
but earn the other scalar-torsion coupling 
$\hat{T}^\rho_{\;\,\rho\mu}\hat{\partial}^\mu \hat{\phi}$ 
at the same time \cite{Yang:2010ji}, 
where $d\hat{\phi} \equiv \sqrt{6}dF/(2F)$.
This scalar-torsion coupling in the new frame
leads to asymmetry field equations \cite{Wu:2011kh}, and
they once again restrict the generation of 
any non-zero scalar-perturbation mode during inflation.

\section*{acknowledgments}

The author thanks Chao-Qiang Geng, Chung-Chi Lee, Tsutomu Kobayashi, Martin Kr\v{s}\v{s}\'{a}k, and James Nester for helpful discussions 
and also acknowledges the hospitality of National Center for Theoretical Sciences (NCTS) on the early development of this work.


\begin{thebibliography}{99}
\bibitem{Indices}
  We use the spacetime indices $\mu, \nu,...$ for the coordinate and 
  $A, B, ...$ for the orthonormal frame. The lower case indices
  $i, j, ...$ are for the spatial coordinate and the lower case indices
  $a, b, ...$ from the beginning of the alphabet are for the spatial
  components of the orthonormal frame.
\bibitem{Sub-horizon}
  In general, perturbation modes that remain well inside the horizon by 
  the end of inflation may survive since the local Lorentz symmetry 
  is effectively restored in the Minkowski limit 
  where $T \rightarrow 0$, see \cite{Ferraro:2014owa}.
\end{thebibliography}
\end{document}